\documentclass[sigconf]{acmart}

\AtBeginDocument{%
  \providecommand\BibTeX{{%
    \normalfont B\kern-0.5em{\scshape i\kern-0.25em b}\kern-0.8em\TeX}}}

\setcopyright{acmcopyright}
\copyrightyear{2023}
\acmYear{2023}
\acmDOI{XXXXXXX.XXXXXXX}

\acmConference[CIKM]{Conference on Information and Knowledge Management}{October 21--25,
  2023}{Birmingham, UK}

\acmPrice{15.00}
\acmISBN{978-1-4503-XXXX-X/18/06}

\usepackage{booktabs}
\usepackage{multirow}
\usepackage{enumitem}
\usepackage[ruled,linesnumbered,vlined]{algorithm2e}
\setlist[itemize]{leftmargin=10pt}
\setlist[enumerate]{leftmargin=15pt}

\begin{document}

\title{SHARK: A Lightweight Model Compression Approach for Large-scale Recommender Systems}

\author{Beichuan Zhang, Chenggen Sun, Jianchao Tan, \\Xinjun Cai, Jun Zhao, Mengqi Miao, Kang Yin, Chengru Song, Na Mou, Yang Song}
\affiliation{%
  \institution{Kuaishou Technology}
  \city{Beijing}
  \country{China}}
\email{{zhangbeichuan, sunchenggen, jianchaotan}@kuaishou.com}
\email{{caixinjun03, zhaojun09, miaomengqi, yinkang, songchengru, mouna, yangsong}@kuaishou.com}

% \author{Submission ID 556}
% \affiliation{%
%   \institution{}
%   \city{}
%   \country{}}
% \email{}

\begin{abstract}
Increasing the size of embedding layers has shown to be effective in improving the performance of recommendation models, yet gradually causing their sizes to exceed terabytes in industrial recommender systems, and hence the increase of computing and storage costs. To save resources while maintaining model performances, we propose SHARK, the model compression practice we have summarized in the recommender system of industrial scenarios. SHARK consists of two main components. First, we use the novel first-order component of Taylor expansion as importance scores to prune the number of embedding tables (feature fields). Second, we introduce a new row-wise quantization method to apply different quantization strategies to each embedding. We conduct extensive experiments on both public and industrial datasets, demonstrating that each component of our proposed SHARK framework outperforms previous approaches. We conduct A/B tests in multiple models on Kuaishou, such as short video, e-commerce, and advertising recommendation models. The results of the online A/B test showed SHARK can effectively reduce the memory footprint of the embedded layer. For the short-video scenarios, the compressed model without any performance drop significantly saves 70\% storage and thousands of machines, improves 30\% queries per second (QPS), and has been deployed to serve hundreds of millions of users and process tens of billions of requests every day.
\end{abstract}

\maketitle
\section{Introduction}
\begin{figure}[!t]
\centering
% % \begin{minipage}[t]{0.48\textwidth}
\centering
\includegraphics[width = 0.4\textwidth]{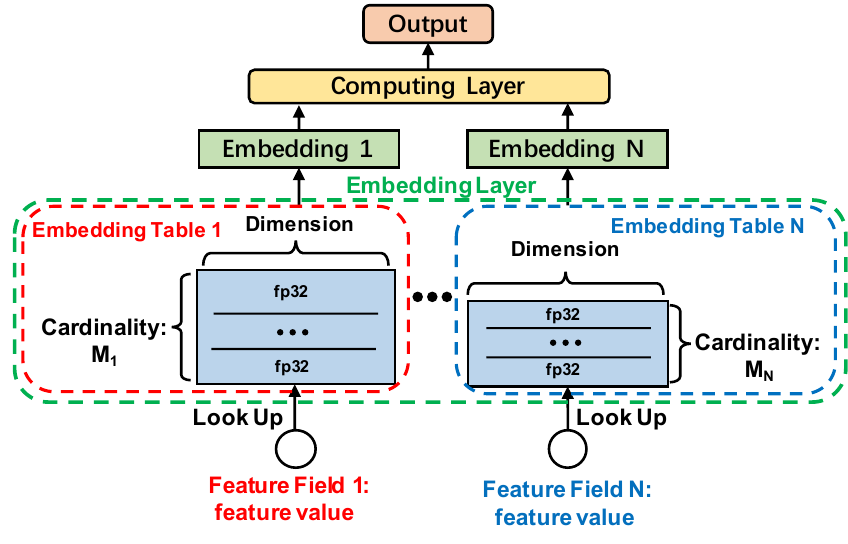}
% \end{minipage}
\small
\caption{A common structural paradigm of recommendation models, consisting of computing layers and embedding layers. An embedding table in the embedding layers corresponds to a feature field, such as age, gender, etc. The number of embeddings in an embedding table is its cardinality. Best viewed in color.
}
\label{fig: structure1}
\vspace{-10pt}
\end{figure}
Building an effective recommender system to recommend accurate and personalized information to users has become a main topic recently thanks to the prevalence of online services such as e-commerce and short-video platforms~\cite{dien, can, dlrm}. Numerous recommendation products across various domains have allocated substantial resources toward enhancing their recommendation models. Model optimization strategies, including feature expansion, embedding dimensionality expansion, feature crossing~\cite{can, DCN}, and long-term interest modeling~\cite{sim1, sim}, have been extensively pursued. However, these approaches often lead to the creation of terabyte-scale embedding layers, which impose a significant burden on the system and hinder the efficiency of model optimization~\cite{COLD, FSCD}. 

To mitigate these problems, it is imperative to employ measures aimed at minimizing the computing and storage resources required by recommendation models, while simultaneously ensuring optimal performance. The compression techniques commonly applied to recommender systems predominantly concentrate on compressing the embedding layers, as evidenced by prior research~\cite{autodim, AutoEmb, ESAPN, nis, LASSO, FSCD, fisher2019shuffle}, which represent a substantial proportion of the overall system. The compression methodologies aim to reduce the memory footprint of the embedding layers through four key dimensions: (1) the quantity of embedding tables~\cite{COLD, FSCD, fisher2019shuffle}; (2) the cardinalities of embedding tables~\cite{Kraken, Monolith, hash, multi-hash}; (3) the dimensions of embeddings~\cite{autodim, PEP, nis}; and (4) the numerical precision~\cite{facebookQuent, facebook_s_q, facebook_frequency_q}, as illustrated in Figure~\ref{fig: structure1}. In this paper, we focus on the \emph{number} of embedding tables and \emph{numerical precision}:
\begin{itemize}
    \item \textbf{Number:} Feature selection methods offer a twofold benefit by reducing the model size and providing insights into the importance-ranking of features in base models. In practical applications, analyzing the significance of feature importance in base models proves valuable for understanding and optimizing the business model. While previous methods have demonstrated efficient feature evaluation capabilities~\cite{COLD, FSCD, LASSO}, they often introduce new structures or parameters into the base model, leading to additional costs and unintended interference. To circumvent these limitations and preserve the structure of existing base models or systems, the industry commonly employs Permutation~\cite{fisher2019shuffle} as a feature evaluation method. However, it is important to note that the computational complexity of Permutation is O($|DATA| \cdot N \cdot T$), where $|DATA|$ denotes the number of data samples, $N$ represents the number of feature fields, and $T$ represents the number of shuffles. This method exhibits inefficiency due to enormous the number of feature fields and the size of the datasets in large-scale recommender systems.
    \item \textbf{Numerical Precision:} In the context of low-precision applications, our investigation reveals that frequently accessed rows (embeddings) are prone to higher errors following quantization.  While previous methods\cite{facebookQuent, facebook_s_q, aac, ALPT} primarily focus on addressing rounding bias and quantization resolution, they often overlook this phenomenon. Another prior approach\cite{facebook_frequency_q} approach~\cite{facebook_frequency_q} makes a similar observation and employs caching techniques to store high-priority embeddings as floating-point precision. However, the priority values in ~\cite{facebook_frequency_q} fail to consider the greater importance of positive samples in the training process. Moreover, previous approaches \cite{facebook_frequency_q, aac, ALPT} ignore the potential advantages of implementing multiple low-precision strategies, such as utilizing both fp16 and int8 simultaneously in the embedding table, as opposed to relying on a single low-precision strategy. This oversight could lead to diminished performance benefits.
\end{itemize}
To address these challenges, we present SHARK, a comprehensive compression methodology that encapsulates the practices derived from industrial recommender systems. SHARK comprises two key components: (1) \textbf{Feature Selection with Fast Permutation (F-Permutation):} F-Permutation is proposed to reduce the number of embedding tables. It leverages the first-order component of the Taylor expansion (Equation~\ref{Taylor1}) as a score for each embedding table, providing an efficient approximation of the original Permutation method~\cite{fisher2019shuffle}. F-Permutation significantly reduces the computational complexity from $O(|DATA| \cdot N \cdot T)$ to $O(|DATA|)$, enabling the inclusion of a larger number of samples for improved feature selection performance. (2) \textbf{Numerical Precision Reduction with Frequency-based Quantization (F-Quantization):} F-Quantization introduces a novel row-wise quantization method. It involves scoring each row in all embedding tables, with different coefficients assigned to positive and negative weights during score updating. F-Quantization enables the embedding layer to operate more efficiently by applying distinct low-precision strategies, such as int8 and fp16, to different embeddings based on their priority values. We conduct extensive experiments to demonstrate that each component of SHARK surpasses previous approaches on both public and industrial datasets. Specifically, this paper makes the following contributions:
\begin{enumerate}
\item Within the framework of SHARK, we introduce F-Permutation, a technique that offers a \emph{significant reduction} in computational cost compared to Permutation~\cite{fisher2019shuffle}. Additionally, we propose F-Quantization, which employs a more efficient priority function and a streamlined framework for applying different low-precision strategies to diverse embeddings. Both methods are well-compatible in industrial recommender system scenarios.
% \item We present that the two approaches are well-compatible in industrial recommender system scenarios.
\item We conduct extensive experiments utilizing both public and industrial datasets to demonstrate that each component of SHARK surpasses previous approaches. In our industrial environment, the compressed model has already served 100\% of users on our application with over 300 million daily active users, resulting in remarkable savings of 70\% in storage capacity and thousands of machines. 
\end{enumerate}

\section{Related Work}
Based on the memory composition, the compression techniques for the embedding layer can be categorized into four primary aspects: the quantity of embedding tables~\cite{COLD, FSCD, fisher2019shuffle}, the cardinalities of embedding tables~\cite{Kraken, Monolith, hash, multi-hash}, the dimensions of embeddings~\cite{autodim, PEP, nis}, and the numerical precision of embeddings~\cite{facebookQuent, facebook_s_q, facebook_frequency_q, ALPT, aac}. In our research, we focus on reducing the number of embedding tables and decreasing the numerical precision of embeddings.
\subsection{Feature Selection}
One common approach to feature selection involves generating a ranking of feature importance and subsequently removing tables with low scores. Previous research has introduced various types of feature selection methods. In the context of traditional recommendation models~\cite{gbdt, FM}, feature selection methods typically assign scores to each feature using predefined metrics, such as correlation coefficients between features and labels. On the other hand, feature selection methods for deep learning-based models have garnered increasing attention from both academia and industry~\cite{L12, Laplacian, IVFS, AutoFAS}. These methods often incorporate sparsity-related constraint losses into the existing model's training procedure to learn the importance scores of feature fields.
COLD~\cite{COLD} employs the squeeze and excitation block~\cite{SEBLOCK} for feature group selection. AutoField and FSCD~\cite{FSCD} utilize the Gumbel-softmax technique for feature selection. Group LASSO~\cite{LASSO}, on the other hand, is a prevalent feature selection method that employs the proximal SGD algorithm~\cite{SSGD} to encourage weight parameters to approach zero. AutoFAS~\cite{AutoFAS} adopts a binarization process for real-valued mask parameters to select significant features. While these methods introduce new structures and parameters into the base model, incurring additional training costs, it is essential in the industry to swiftly provide researchers with feature importance analysis for the original base models.

In addition to these training-based methods, the Permutation~\cite{fisher2019shuffle} technique is widely employed in industry, specifically during the model inference stage. Permutation determines the importance score of feature fields by evaluating performance degradation following the permutation of embeddings within a batch. However, the Permutation-based method exhibits high time complexity when dealing with billions of data samples.

\subsection{Embedding Table Quantization}
Quantization is a highly effective technique for reducing the memory footprint of models. In the context of recommendation models, quantization methods primarily focus on the compression of embedding layers~\cite{facebookQuent, facebook_s_q, aac, ALPT, facebook_frequency_q}. The quantization methods applied to embedding tables in recommender systems can be categorized into two main types: post-training quantization and training-aware quantization. In this study, our emphasis lies on the latter.

One approach for post-training quantization involves row-wise quantization using greedy search and codebook-based techniques, as proposed in~\cite{post-q}. On the other hand, various quantization algorithms aim to compress embedding tables during the training process. Deng et al. tailor their low-precision strategy to adapt the recommendation model to low-precision hardware~\cite{facebookQuent}. Zhang et al. introduce stochastic floating-point rounding to achieve more effective quantization~\cite{facebook_s_q}. ALPT~\cite{ALPT} proposes an adaptive low-precision training framework to learn quantization scales. However, these approaches fail to address the phenomenon we have discovered, wherein frequently accessed rows (embeddings) experience higher errors after quantization.

Yang et al. \cite{facebook_frequency_q} also observe a similar phenomenon and propose the use of Least Frequently Used (LFU) or Least Recently Used (LRU) policies to determine which embeddings are stored in a full-precision cache~\cite{facebook_frequency_q}. Nevertheless, the priority values derived from LFU or LRU overlook the importance of positive samples in the training process. Moreover, the cache architecture presents challenges in terms of scalability when multiple low-precision strategies (such as fp16, int8, etc.) need to be simultaneously applied, thereby impeding the optimization of embedding tables' efficiency.
\section{Method}
\subsection{F-Permutation}
\label{sec: F-Permutation}
% Figure \ref{fig: pfstructure} illustrates the process of F-Permutation, which consists of three steps. In step 1, F-Permutation applies Equation \ref{Taylor1} to assign a score to each embedding table within a recommendation model. In step 2, F-Permutation proceeds to remove embedding tables with the lowest scores and subsequently fine-tunes the pruned model accordingly. Finally, in step 3, the pruned model is executed on test data to compute gradients and evaluate the results.
\subsubsection{Table-wise Scores Calculating}
In the process of model convergence, it becomes evident that different feature fields contribute to accuracy to varying degrees. Consequently, it becomes imperative to identify unimportant feature fields and eliminate the corresponding embedding tables to minimize any loss in accuracy. Permutation \cite{fisher2019shuffle} introduces a method to assess the performance degradation of a sample $x$ by replacing the original candidates with candidates from other samples while keeping the remaining feature fields unchanged. If replacing the values of a feature field significantly increases the model error, it is considered important. Conversely, if shuffling the values of a feature field leaves the model error unchanged, it is deemed unimportant. Formally, we define the prediction error of the $i_{th}$ feature field in the sample $x$ after shuffling as:
\begin{equation}
\small
\begin{split}
error(i,x) = \sum_{ v_i' \in set_i }& loss(v_1^*,...,{v_i'},...,v_N^*)p({v_i'})-\\
                                   & loss(v_1^*,...,{v_i^*},...,v_N^*)\\
\end{split}
\label{importance1}
\end{equation}
where $N$ represents the total number of feature fields, ${v_i'}$ represents arbitrarily candidates in $set_i$, $set_i$ contains all feature value in the $i_{th}$ feature field, $v_n^*$ represents the original value of $n_{th}$ feature field in sample $x$, and $p({v_i'})$ represents probability of the value ${v_i'}$ in whole dataset. The prediction error corresponding to the $i_{th}$ feature field in the whole dataset after shuffling is defined as:
\begin{equation}
\small
\begin{split}
error(i) = \frac{1}{|DATA|} \sum_{ {x} \in {DATA} } error(i,x)
\end{split}
\end{equation}
where $|DATA|$ represents the number of samples in the dataset $DATA$. The larger the value of $error(i)$, the more important the $i_{th}$ feature field is.  The above equation described the evaluation method for one feature field. Furthermore, the list of table-wise scores for all feature fields $W_t$ is defined:
\begin{equation}
\small
\begin{split}
W_t = \{w_t^1=error(1),...,w_t^N=error(N)\}
\end{split}
\end{equation}
Consequently, the time complexity involved in calculating $W_t$ becomes O($|DATA| \cdot N \cdot \bar{|c|}$), where $\bar{|c|}$ denotes the average cardinality of all feature fields. This computation becomes prohibitively expensive due to the enormous average cardinality of feature fields and the size of the datasets. In industrial settings, the Permutation method relies on shuffling operations performed $T$ times to approximate $W_t$, which still incurs a high time complexity. To address this issue, we propose the utilization of Taylor expansion to approximate $W_t$. Specifically, we employ the first-order Taylor expansion of Equation \ref{importance1}, as demonstrated in Equation \ref{Taylor1}.
\begin{equation}
\small
\begin{split}
error(i,x) &= \frac{ \partial loss(v_1^*,...,{v_i^*},...,v_N^*) }{ \partial {v_i^*} }\sum_{ v_i' \in set_i }p({v_i'})(v_i'-{v_i^*})\\
       &= \frac{ \partial loss(v_1^*,...,{v_i^*},...,v_N^*) }{ \partial {v_i^*} }(E(v_i)-{v_i^*})
\end{split}
\label{Taylor1}
\end{equation}
where $E(v_i)$ stands for expectations of the $i_{th}$ feature field. 
Following the approximation, the time complexity for calculating $W_t$ becomes $O(3 \cdot |DATA|)$. The time complexity for computing the average value of all features in advance is $O(|DATA|)$, while the time complexity for forward and backward propagation is $O(2 \cdot |DATA|)$. We have tried using the second-order Taylor expansion of Equation \ref{importance1}. The experimental results show the performance of the first-order variant is similar to that of the second-order variant. To reduce computational complexity, we use the first-order variant.
\subsubsection{Model Pruning and Fine-tuning}

As the information provided by feature fields can be interconnected, the importance of existing feature fields may change after deleting certain feature fields \cite{fisher2019shuffle}. To mitigate the impact of information coupling, F-Permutation adopts an iterative approach to remove unimportant embedding tables. Specifically, after calculating the table-wise scores $W_t$, F-Permutation selectively eliminates a predetermined number of $f$ embedding tables with the lowest scores (with a default value of 1). However, this removal of embedding tables disrupts the convergence of the model. Rebuilding a model from random initialization is a time-consuming process. Due to the minor changes made, we can utilize a small subset of data to retrain the model and achieve convergence once again.
\subsubsection{Pruned Model Evaluation}
We evaluate the pruned model on a test dataset using machine learning frameworks. To determine the termination criterion for the compression process, we introduce two thresholds denoted as $rate_{c}$ and $T_{accuracy}$. The compression process halts when either the memory cost falls below the $rate_{c}$ threshold or the performance of the pruned model falls below the $T_{accuracy}$. In our production environment, we consider an accuracy drop exceeding 0.15\% to be significant. Therefore, we set $T_{accuracy}$ as $99.25\% $ the accuracy of the base model to monitor the performance degradation during compression.
\subsubsection{Whole Pipeline}
We summarize the aforementioned steps in algorithm~\ref{algorithm1}.
\begin{algorithm}[!hbt]
\footnotesize
\caption{Fast Permutation}
\label{algorithm1}
\KwIn{Convergent Model, $M$; Target Compression Ratio, $rate_{c}$;}
\KwData{Traing Dataset, $DATA_{train}$; Test Dataset, $DATA_{test}$;}
$|DATA_{test}| \leftarrow $ The number of samples in $DATA_{test}$\;
\tcp{$LookUp(DATA_{test})$ can select the embeddings according to $|DATA_{test}|$}
expectation $\leftarrow LookUp(DATA_{test}) / |DATA_{test}|$\;
\tcp{$rate_{t}$ is current compression rate}
$rate_{t} \leftarrow 0$\;
performance, gradients $\leftarrow$ Evaluation($M$, $DATA_{test}$)\;
\While{$rate_{t} \leq rate_{c}$ and $T_{accuracy} \leq$ performance }{
    TableWiseScores $\leftarrow$ Equation~\ref{Taylor1} (gradients, expectation)\;
    \tcp{delete tables with lowest scores}
    PrunedModel, $rate_{t} \leftarrow$ DeleteTables (TableWiseScores)\;
    \tcp{FinetuneModel represents models that are trained on a small amount of data.}
    $DATA_{support} \leftarrow$ sample ($DATA_{train}$)\;
    M $\leftarrow$ FinetuneModel (PrunedModel, $DATA_{support}$)\;
    performance, gradients $\leftarrow$ Evaluation ($M$, 
 $DATA_{test}$)\;

}
\KwOut{M}
\end{algorithm}
\vspace{-10pt}

\subsection{F-Quantization}
In this section, we present our low-precision strategy for embedding during training. The quantization and dequantization equations for embedding $e$ are as follows:
\begin{equation}
\small
\begin{aligned}
e_{q} &=\mathcal{Q}\text{uantization}(e)= \operatorname{round}(\frac{e}{\mathrm{scale}}) \\
e_{dq}&=\mathcal{D}\text{equantization}\left(e_{q}\right)=\text { scale } \cdot e_{q}
\end{aligned}
\label{quantization}
\end{equation}
Large-scale recommendation models typically comprise billions of parameters, with parameter values exhibiting a wide distribution. Applying a single scale to all embedding tables or assigning a scale to each individual embedding table can result in unacceptable quantization errors~\cite{facebookQuent}. To mitigate this issue, we employ row-wise quantization, whereby a distinct scale is assigned to each row within the embedding tables.
\begin{equation}
\small
% \begin{aligned}
\text { scale }=\frac{e^{abs}_{max}}{I_{\max }-I_{\min }},
% \end{aligned}
\label{scale}
\end{equation}
where the value $e^{abs}_{max}$ represents the maximum absolute value found in the embedding vector $e$. For the target $b$-bit integer type, the range $[I_{\min}, I_{\max}]$ is defined as $[-2^{(b-1)}, 2^{(b-1)}-1]$. However, in our recommender system, we observed significant quantization errors when employing row-wise quantization with fp16 precision. Through analysis, we discovered that frequently accessed rows tend to exhibit higher quantization errors, as reported in~\cite{facebook_frequency_q}. To address this issue, firstly,  we define the  frequency-based priority value  $\mathbf{w}_{\mathrm{r}}$ of row $r$ at time $t$ as follows:
\begin{equation}
\small
\begin{split}
\mathbf{w}_{\mathrm{r}}^{(t+1)}=(1-\beta) \mathbf{w}_{\mathrm{r}}^{(t)}+\beta\left(c^{+} \alpha +c^{-}\right)
\end{split}
\label{importance embedding}
\end{equation}
where $\beta$ is the time decay rate, $\alpha$ is the importance weight, $c^{+}$ and $c^{-}$ are the number of positive and negative examples containing row $r$ in a batch, respectively. Since positive samples are more important in the training process, we assign different coefficients to positive and negative weights. The score $\mathbf{w}_{\mathrm{r}}$ of the embedding $r$ will be updated per batch. In our method, the rows with the highest, middle, and lowest scores can be trained in full precision, fp16 precision, and int8 precision, as the following equation:
\begin{equation}
\small
\begin{aligned}
   r_{q}&= \mathcal{Q}(r)=
    \left\{
        \begin{aligned}
            & \operatorname{rnd_{8}}(\frac{r}{\mathrm{scale_{int8}}}) & \text{if } \mathbf{w}_{\mathrm{r}} < t_{8}\\
            & \operatorname{rnd_{16}}(\frac{r}{\mathrm{scale_{fp16}}}) & \text{if } t_{8} \leq \mathbf{w}_{\mathrm{r}} < t_{16}\\
            & r & \text{if } t_{16} \leq \mathbf{w}_{\mathrm{r}}\\
        \end{aligned}
    \right.
% \end{equation}
% \begin{equation}
\\\\
  r_{dq}&= \mathcal{D}\left(r_{q}\right)=
    \left\{
        \begin{aligned}
          & r_q \cdot \mathrm{scale_{int8}} & \text{if } \mathbf{w}_{\mathrm{r}} < t_{8}\\
            & r_q \cdot \mathrm{scale_{fp16}} & \text{if } t_{8} \leq \mathbf{w}_{\mathrm{r}} < t_{16}\\
            & r_q & \text{if } t_{16} \leq \mathbf{w}_{\mathrm{r}}\\
        \end{aligned}
    \right.
\end{aligned}
\label{h_q}
\end{equation}
The symbol $rnd_{8}$ denotes a low-precision strategy using int8 quantization, while $rnd_{16}$ represents a low-precision strategy using fp16 quantization. The vector $\mathbf{w}_{\mathrm{r}}$ corresponds to the row-wise scores of the embedding $e$. The hyperparameters $t{16}$ and $t_{8}$ are used in the quantization process, and their suitable values are determined by evaluating the prediction error of compressed models after applying the respective quantization strategy to the embedding tables. The search for appropriate values of $t_{16}$ and $t_{8}$ is performed independently, taking into consideration the specific requirements of each quantization strategy. Extra words in each embedding have been applied to handle the conflict of the universal description for different precisions as shown in Table \ref{byte}.
\begin{table}[]
% \footnotesize
\small
\caption{Byte layout of extra words.}
\label{byte}
\begin{tabular}{@{}ccccc@{}}
\toprule
Name & Precision & Dimension & Scale \\ \midrule
Byte & 8bit     & 16bit     & 32bit \\ \bottomrule
\end{tabular}
\vspace{-10pt}
\end{table}
\section{Experiments}
\subsection{Experimental Settings}
\label{sec: Experimental Settings}
\subsubsection{Dataset}
To validate the efficacy of our method, we conducted experiments on the Criteo dataset\footnote{https://labs.criteo.com/2013/12/download-terabyte-click-logs}, and the industrial dataset from Kuaishou. The Criteo dataset consists of 26 categorical feature fields and 13 continuous feature fields, encompassing approximately 4.4 billion click samples recorded over a span of 24 days. Following the setting of DLRM~\cite{dlrm}, we randomly selected 12.5\% of the samples per day to serve as negative samples. For the Criteo dataset, we divided the data into two parts: data from days 0 to 17 and data from days 18 to 22. The former was used for pre-training and compression of the model described in Section \ref{sec: Model Settings}. Subsequently, after compression, we employed data from days 0 to 22 for model retraining and evaluation. We conducted an A/B test wherein both the baseline model and the compressed model were subjected to 10\% of the online traffic. The A/B test was conducted over a period of 7 days. We need to save more than 40\% of the memory footprint while the accuracy drop is less than 0.15\%, according to practical experience in the industry.
\subsubsection{Model Settings}
\label{sec: Model Settings}
In our experiments, we utilized the DLRM~\cite{dlrm} model as the baseline for the Criteo dataset. The dataset consists of 13 continuous features, which were transformed into embeddings, along with 26 discrete feature fields, resulting in a total of 27 feature fields. Following the approach employed in UMEC~\cite{UMEC}, we set the hidden dimensions of the three-layer MLP prediction model to 256 and 128. 
For the industrial dataset, we employed the master ranking model specifically designed for our short-video recommendation scenario. This model comprises 180 features and tens of billions of parameters. It is a multi-task model, catering to various objectives such as clicks, likes, follows, and more, based on the approach described in MMOE~\cite{mmoe}.

In the case of the Criteo dataset, we used a batch size of 512 and a learning rate of 0.01. As for the industrial dataset, the batch size was set to 8192, with a learning rate of 0.01. In the case of competitor methods, we determined the optimal learning rate from \{0.01, 0.005, 0.001, 0.0005, 0.0001\}, and the coefficient $\lambda$ for L2 regularization was selected from \{${10^{-4},10^{-5},10^{-6},10^{-7},10^{-8}}$\}. $\alpha$ and $\beta$ were set to 2 and 0.99. We conducted a comparative evaluation of our proposed methods and other approaches using an online learning methodology on all the datasets.
\begin{figure}[!t]
\centering
\includegraphics[width = 0.35\textwidth]{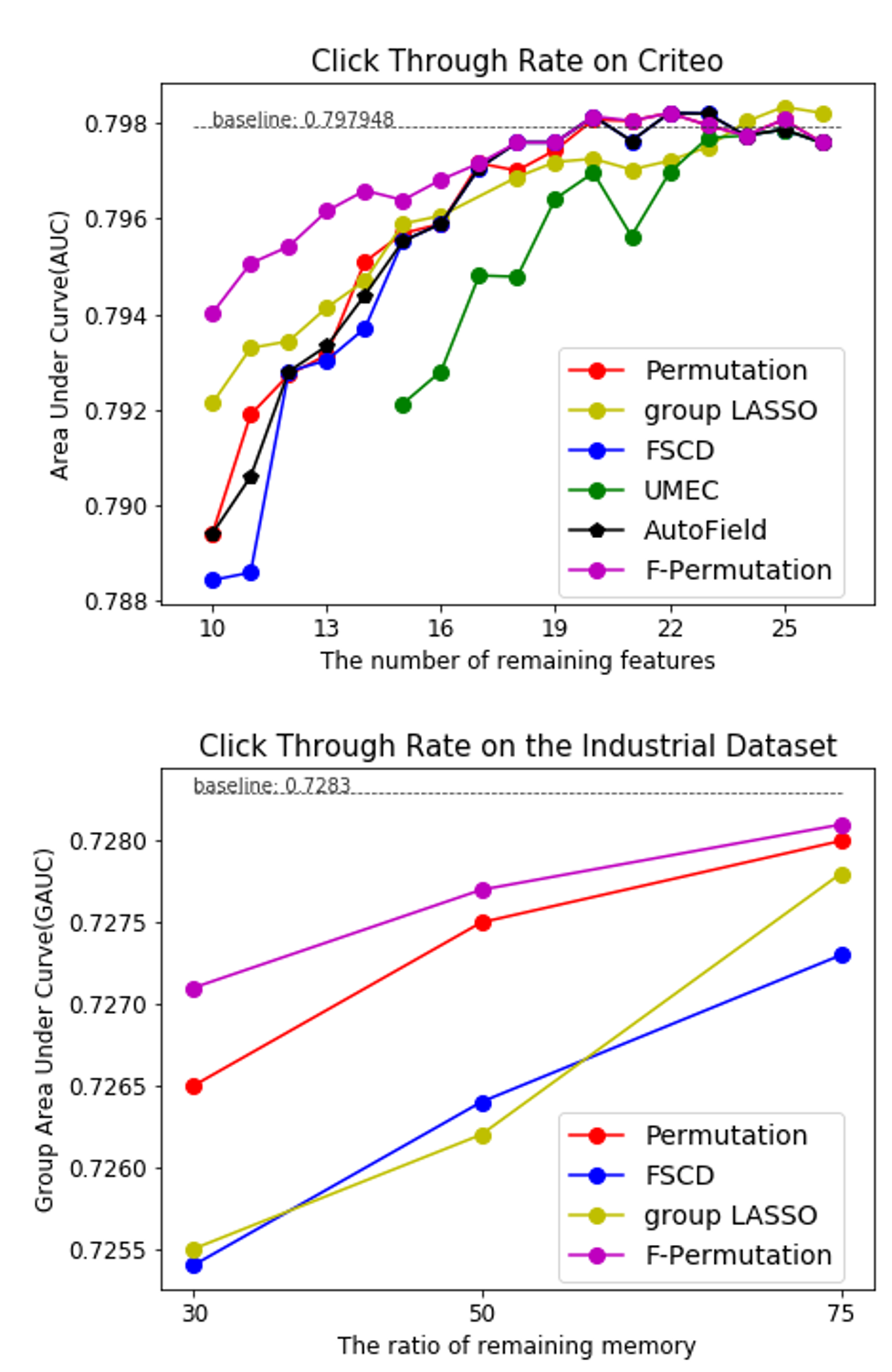}
\caption{Performance comparison from the Criteo dataset and industrial dataset. The line named "baseline" represents the performance of the model without any pruning.}
\label{fig: fpCriteo}
\vspace{-12pt}
\end{figure}

% \begin{figure}
% \centering
% \includegraphics[width = 0.35\textwidth]{pffig/fpCTR.png}
% \caption{Performance comparison on the industrial dataset. The line named "baseline" represents the performance of the model without any pruning.}
% \label{fig: fpCTR}
% \vspace{-12pt}
% \end{figure}

\subsubsection{Baselines}
% We compare our model with the following baselines:
\begin{itemize}
\item{\textbf{FSCD}~\cite{FSCD}}: FSCD learns the importance scores of feature fields by the Gumbel-softmax technique and then sets a threshold to remove unimportant fields.
\item{\textbf{AutoField}~\cite{AutoField}}: AutoField uses the Gumbel-softmax technique and a bi-level optimization for feature selection.
\item{\textbf{group LASSO}~\cite{LASSO}}: With the proximal SGD\cite{proximalSGD}, LASSO can push a portion of parameters to be exact zeros. We regularize the weights which directly connect with the output of the embedding layer and train them by proximal-SGD.
\item{\textbf{UMEC}~\cite{UMEC}}: UMEC uses an alternating direction method of multipliers optimization to select features. 
\item{\textbf{Permutation}~\cite{fisher2019shuffle}}: This method shuffles the values of a feature field in a batch and keeps other fields unchanged. To reduce the computational complexity, we only randomly shuffle once.
\item{\textbf{Mixed-Precision Embedding (MPE)}~\cite{facebook_frequency_q}}: This method uses a floating point precision cache with LFU or LRU to achieve low-precision.
\item{\textbf{ALPT}~\cite{ALPT}}: ALPT proposes an adaptive low-precision training framework to learn quantization scale.
\item{\textbf{PEP}~\cite{PEP}}: PEP proposes to use learnable threshold(s) to determine the dimensions of embeddings. We use the feature-dimension variant of PEP, the finest granularity~\cite{PEP},  to search the dimension of each embedding.
\end{itemize}
\subsection{Performance of F-Permutation}
\label{sec: Performance of F-Permutation}
Figure~\ref{fig: fpCriteo} shows the performance of all methods on Criteo and industrial datasets. Table~\ref{tab: Time analysis} shows the time analysis of all methods on the industrial dataset. Our main observations are as follows:
\begin{table}[]
\footnotesize
\caption{Time analysis of all methods on the industrial dataset. 
% Bold scores are the best. 
"Model pruning" refers to the total time of the pruning process. "Scores Producing" refers to the time of table-wise score producing in the model pruning process. }
\label{tab: Time analysis}
\begin{tabular}{@{}ccccc@{}}
\toprule
Method           & FSCD~\cite{FSCD}  & LASSO~\cite{LASSO} & Permutation~\cite{fisher2019shuffle} & \textbf{F-P}    \\ \midrule
Scores Producing & 3 days & 3 days & 6 hours      & \textbf{1 hour} \\
Model Pruning    & 3 days & 3 days & 2 days       & \textbf{1 day}  \\ \bottomrule
\end{tabular}
\end{table}

\begin{enumerate}
\item  F-Permutation achieves the best results in all datasets, with the exception when the number of remaining features reaches 22-26 in the Criteo dataset. Our method does not introduce additional hyper-parameters or learnable parameters, thus being more robust to compress the existing models with minimal performance drop, especially in large-scale industrial models.
\item Compared with the original Permutation-based method, ours achieves better results. The complexity of the original Permutation-based method can only support random shuffling in a mini-batch and is not efficient when data samples are billions-scale, thus ignoring the overall properties of whole data and falling into local optimum.
\item As shown in Figure~\ref{fig: fpCriteo}, when only a few features are removed (i.e. the remaining feature domains are more than 19), the accuracy loss is almost zero. One explanation is that the information from different feature fields is highly coupled before compression, therefore, evicting some features is lossless.
\item Table~\ref{tab: Time analysis} shows that compared with other methods, our method has a significant time promotion, because it does not involve training new model parameters, changing model structure, etc.

% As shown in Table~\ref{tab: Time analysis}, 
\end{enumerate}

\begin{table}[]
% \footnotesize
\small
\caption{Performance comparison of our F-Quantization(F-Q) with other methods on the industrial dataset. we set $t_8$ as \textbf{1e3} and $t_{16}$ as \textbf{1e5}. Bold scores are the best. Accuracy Drops greater than 0.15\% are significant in our production.}
\label{tb: fqoffline}
\centering
\begin{tabular}{@{}ccccc@{}}
\toprule
       & FP32   & \textbf{F-Q}                                                        & MPE~\cite{facebook_frequency_q}                                                         & ALPT~\cite{ALPT}                                                        \\ \midrule
Click  & 0.7283 & \textbf{\begin{tabular}[c]{@{}c@{}}0.7293\\  (0.13\%)\end{tabular}} & \begin{tabular}[c]{@{}c@{}}0.7255\\  (-0.38\%)\end{tabular} & \begin{tabular}[c]{@{}c@{}}0.7244 \\ (-0.54\%)\end{tabular} \\
Like   & 0.7496 & \textbf{\begin{tabular}[c]{@{}c@{}}0.7509\\ (0.17\%)\end{tabular}}  & \begin{tabular}[c]{@{}c@{}}0.7457\\ (-0.52\%)\end{tabular}  & \begin{tabular}[c]{@{}c@{}}0.7433 \\ (-0.84\%)\end{tabular} \\
Follow & 0.7938 & \textbf{\begin{tabular}[c]{@{}c@{}}0.7928\\ (-0.12\%)\end{tabular}} & \begin{tabular}[c]{@{}c@{}}0.791\\  (-0.35\%)\end{tabular}  & \begin{tabular}[c]{@{}c@{}}0.7886 \\ (-0.65\%)\end{tabular} \\
Memory & 100\%  & \textbf{50\%}                                                       & 55\%                                                        & 55\%                                                        \\ \bottomrule
\end{tabular}
\vspace{-3pt}
\end{table}

\begin{figure}[]
\centering
\includegraphics[width = 0.35\textwidth]{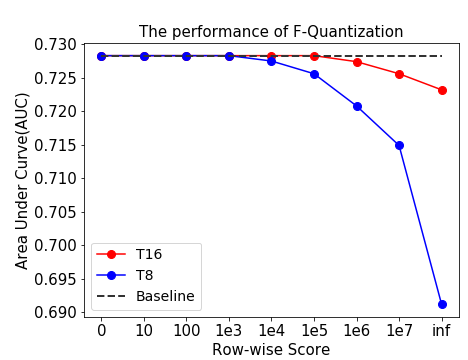}
\caption{Performance of F-Quantization w.r.t different  $t_{16}$ and $t_{8}$. In the figure, "T8" and "T16" represents $t_{8}$ and $t_{16}$. "Baseline" represents the performance of the model with fp32. Best viewed in color.}
\label{fig: ablation_ctr}
% \vspace{-9pt}
\end{figure}
% \begin{figure}[!t]
% \centering
% \centering
% \includegraphics[width = 0.42\textwidth]{pffig/fq_m_ablation.png}
% \caption{Memory footprint of embedding layers w.r.t different  $t_{16}$ and $t_{8}$. In the figure, "T8" and "T16" represents $t_{8}$ and $t_{16}$. "Baseline" represents the performance of the model with fp32. When we show the results of F-Quantization about different $t_8$, we set $t_{16}$ to the same value as $t_{8}$. When we show the results of F-Quantization about different $t_{16}$, we set $t_{8}$ to zero. Best viewed in color.}
% \label{fig: ablation_me}
% \end{figure}

\subsection{Performance of F-Quantization}
\label{sec: Performance of F-Quantization}
Table~\ref{tb: fqoffline} shows the performance of F-Quantization and related methods in our asynchronous training frameworks on the industrial dataset. We perform ablation studies on F-Quantization to show how the $t_{16}$ and $t_{8}$ affect performance in Figure~\ref{fig: ablation_ctr}.
% Figure~\ref{fig: ablation_ctr} and Figure~\ref{fig: ablation_me} shows the results of F-Quantization w.r.t different $t_{16}$ and $t_{8}$. 
When we show the results of F-Quantization about different $t_8$, we set $t_{16}$ to the same value as $t_{8}$. When we show the results of F-Quantization about different $t_{16}$, we set $t_{8}$ to zero.
\begin{enumerate}
\item Compared with fp32, our method reduces total memory usage by 50\% and achieves the same performance. 
Compared with fp16 with stochastic rounding, at 55\% memory footprint, our method achieves better performance about AUC on all prediction tasks (such as Like, Click, and Follow), which may demonstrate the effectiveness of F-Quantization. 
\item Training with int8 stochastic rounding decreases more than $2\%$ AUCs on all tasks, which is a significant drop in our industrial productions.
\item Figure~\ref{fig: ablation_ctr} shows when $t_{16}$ is less than \textbf{1e5}, the model compressed by F-Quantization can maintain the performance achieved by the model with fp32. When $t_8$ is less than \textbf{1e3}, the model compressed by F-Quantization can maintain the performance achieved by the model with fp32. In conclusion, the best $t_8$ is \textbf{1e3}, and the best $t_{16}$ is \textbf{1e5}.
\end{enumerate}

\subsection{Combined Performance of F-Permutation \& F-Quantization}
Table~\ref{tb: sharkp} illustrates the combined performance of two methods, compared with individual ones. We observe that:
\begin{enumerate}
    \item F-Quantization and F-Permutation can reduce the memory footprint to 50\% and 60\%, respectively, with a negligible performance drop compared to the baseline. Our methods achieve notable improvements while maintaining the same memory usage as PEP.
    \item We directly merge the compression results of F-Quantization and F-Permutation to show their compatibility. The combination of them ( F-Q + F-P ) successfully reduces the memory footprint to 30 (50 $\times$ 60)\% of the baseline methods, while maintaining competitive performance as its components. This result suggests that F-P and F-Quantization are highly compatible with each other.
\end{enumerate}

\begin{table}[!t]
\caption{Performance of combinations of F-Permutation and F-Quantization on the industrial dataset. F-Q and F-P are short for F-Quantization and F-Permutation. }
\label{tb: sharkp}
\centering
\small
% \footnotesize
\begin{tabular}{@{}ccccccc@{}}
\toprule
       & Baseline & PEP~\cite{PEP}                                                      & F-Q                                                      & F-P                                                      & PEP~\cite{PEP}                                                      & F-Q+F-P                                                  \\ \midrule
Click  & 0.7283   & \begin{tabular}[c]{@{}c@{}}0.7255\\ -0.39\%\end{tabular} & \begin{tabular}[c]{@{}c@{}}0.7292\\ +0.13\%\end{tabular} & \begin{tabular}[c]{@{}c@{}}0.7277\\ -0.08\%\end{tabular} & \begin{tabular}[c]{@{}c@{}}0.7236\\ -0.64\%\end{tabular} & \begin{tabular}[c]{@{}c@{}}0.7279\\ -0.05\%\end{tabular} \\
Like   & 0.7496   & \begin{tabular}[c]{@{}c@{}}0.7468\\ -0.38\%\end{tabular} & \begin{tabular}[c]{@{}c@{}}0.7509\\ +0.17\%\end{tabular} & \begin{tabular}[c]{@{}c@{}}0.7493\\ -0.04\%\end{tabular} & \begin{tabular}[c]{@{}c@{}}0.7433\\ -0.84\%\end{tabular} & \begin{tabular}[c]{@{}c@{}}0.7493\\ -0.04\%\end{tabular} \\
Follow & 0.7938   & \begin{tabular}[c]{@{}c@{}}0.7912\\ -0.33\%\end{tabular} & \begin{tabular}[c]{@{}c@{}}0.7928\\ -0.12\%\end{tabular} & \begin{tabular}[c]{@{}c@{}}0.7928\\ -0.12\%\end{tabular} & \begin{tabular}[c]{@{}c@{}}0.7886\\ -0.65\%\end{tabular} & \begin{tabular}[c]{@{}c@{}}0.7941\\ +0.04\%\end{tabular} \\
Memory & 100\%    & 60\%                                                     & 50\%                                                     & 60\%                                                     & 30\%                                                     & 30\%                                                     \\ \bottomrule
\end{tabular}
% \vspace{-2pt}
\end{table}

% It is noteworthy that our platform boasts a substantial user base, with a Daily Active Users (DAU) count exceeding 300 million. Therefore, deploying a model that serves 10\% of the traffic translates to catering to tens of millions of users and processing billions of requests on a daily basis. In our offline testing, we considered AUC drops greater than 0.15\% as significant, while in online testing, watch time drops greater than 0.1\% was deemed significant for evaluating the performance of the models.
\section{Conclusion}
In this paper, we proposed SHARK, a lightweight compressing method for embedding layers. We addressed two aspects of redundancies in the embedding layers, namely the number of embedding tables and the numerical precision of embeddings, by introducing two novel methods. Our extensive experiments conducted on various benchmarks and datasets demonstrated the superiority of these methods over previous approaches. In industrial applications, SHARK was successfully deployed in the short video, e-commerce, and advertising recommendation models on our platform. 

Regarding the short video ranking model, the results of the online A/B test showed that compressing 70\% of the storage used by the industrial model did not lead to a decrease in average watch time. Additionally, there was a 30\% improvement in queries per second (QPS), resulting in substantial savings in terms of computational resources. Thousands of machines were saved as a result of these efficiency gains. 

We acknowledge that while SHARK effectively reduced the memory size of the industrial-scale recommendation model using a total compression ratio, determining the optimal partial compression ratio for each component remains labor intensive: we assigned multiple compression ratios to each part and exhaustively explored all combinations to find the optimal solution. Our experience indicated that an even distribution of compression ratios yielded the best outcomes. Automating this above step remains our future work. 
% We expect to extend F-Permutation to reduce the dimension of embedding tables while maintaining the model performance. 
% We will calculate a dimension-wise importance score to adaptively find a suitable dimension for each embedding table. 

%We will apply meta-learning as MAML\cite{maml} to restore the value of evicted embeddings as soon as possible when the evicted embeddings are admitted again.
% We may also assign different precision and cache policies for each embedding based on a lightweight profile in the first few training iterations for the best accuracy. 

\clearpage
\bibliographystyle{ACM-Reference-Format}
\bibliography{./bibliography/ref}

\end{document}